\documentclass[pra,twocolumn,superscriptaddress]{revtex4}  % for review and submission
\usepackage[usenames,dvipsnames]{color}
\usepackage{graphicx}% Include figure files
\usepackage{dcolumn}% Align table columns on decimal point
\usepackage{bm}% bold math
\usepackage{color}
\usepackage{amsmath}
\usepackage{graphicx}
\usepackage{dsfont}

\begin{document}

\title{Cooling a Mechanical Resonator to Quantum Regime by heating it}

\author{Yue Ma}
\affiliation{Center for Quantum Information, Institute for
  Interdisciplinary Information Sciences, Tsinghua University, Beijing
  100084, China}

\affiliation{Department of Physics, Tsinghua University, Beijing
  100084, China}

\author{Zhang-qi Yin}\email{yinzhangqi@mail.tsinghua.edu.cn}
\affiliation{Center for Quantum Information, Institute for
  Interdisciplinary Information Sciences, Tsinghua University, Beijing
  100084, China}

\author{Pu Huang} \affiliation{Hefei National
  Laboratory for Physics Sciences at Microscale and Department of
  Modern Physics, University of Science and Technology of China, Hefei
  230026, China}
\author{W. L. Yang} \affiliation{State Key Laboratory
  of Magnetic Resonance and Atomic and Molecular Physics, Wuhan
  Institute of Physics and Mathematics, Chinese Academy of Sciences,
  Wuhan 430071, China}
\author{Jiangfeng Du}\email{djf@ustc.edu.cn}
\affiliation{Hefei National Laboratory for Physics Sciences at
  Microscale and Department of Modern Physics, University of Science
  and Technology of China, Hefei 230026, China}

\begin{abstract}

We consider a mechanical resonator made of diamond, which contains a nitrogen-vacancy center (NV center) locating at the end of the oscillator. A second order magnetic gradient is applied and inducing coupling between mechanical modes and the NV center.
 By applying proper external magnetic field, the energy difference between NV center electron spin levels
can be tuned to match the energy difference between two mechanical modes $a$ and $b$.
A laser is used for continuously initializing the NV center electron spin.
  The mode $a$ with lower frequency is driven by a thermal bath. We find that the temperature of the mode $b$ is significantly cooled when the heating bath temperature is increased. We discuss the conditions that quantum regime cooling requires, and confirm the results by numerical simulation. Finally we give the intuitive physical explanation on this unusual effect.
\end{abstract}

\pacs{******} 

\maketitle

%%%%%%%%%%%%%%%%%%%%%%%%%%%%%%%%%%%%%%%%%%%%%%%%%%%%%
%\section{Introduction}
%%%%%%%%%%%%%%%%%%%%%%%%%%%%%%%%%%%%%%%%%%%%%%%%%%%%%%
%

\section{introduction}

Cooling and manipulating the motion of mechanical resonator is widely studied now \cite{ASP2014}. It has many applications
in quantum information science \cite{Regal2011,Barzanjeh2012,Wang2012,Tian2012,Rips2013,Yin2009a,Yin2015a,Barzanjeh2015}, testing quantum effects in
macroscopic systems \cite{PEN1996,Connell2010,Teufel2011,Poot2012,Pikovski2012,NIM2013,Bassi2013,Isart2011,Yin2013,Li2016,Barzanjeh2016}, ultra-sensitive
sensing \cite{Caves1981,Geraci2010,Yin2011,Arvanitaki2013,Huang2013,Zhao2014}, etc. In order to cool mechanical resonator, we should couple it to other cold systems such
as cavity mode (optomechanics) \cite{sideband1,sideband2,Liu2013}, electronic spin (Nitrogen-vacancy centers) \cite{RAB2009,MacQuarrie2016,Li2016b}, etc. In optomechanics, laser cooling of mechanical resonators is a typical method. Intuitively, this is not hard to understand. A laser produces coherent
light, which is well ¡°ordered¡± (cold), and will make the mechanical system couple with a colder cavity bath \cite{sideband1,sideband2}.
However, counter-intuitively, recent research showed that a thermal light, which is ¡°very hot¡±, can also cool
an opto-mechanical system~\cite{MAR2012}. In this system, the mechanical resonator to be cooled is coupled with
two optical modes \cite{Yin2009}. The low frequency mode is in contact with a hot thermal light while the high frequency mode
is not, acting as an auxiliary part. After the system reaches the equilibrium state, the thermal phonon
number in the mechanical mode will be smaller.

In a hybrid system that contains a NV center and a mechanical resonator, a magnetic gradient is applied
to couple the NV center with the resonator. The NV center is continuously initialized by a laser, and could be used for
cooling the mechanical resonator to the ground state \cite{RAB2009}.
In previous literatures, the first order magnetic gradient was used for inducing coupling between the NV center and the mechanical
mode \cite{RAB2009,YIN2015,Zhou2010,Xu2009,Chen2010,Arcizet2011,Kolkowitz2012}.
It is quite naturally to ask whether the higher order magnetic gradient could
be useful for coupling the NV center and the mechanical resonator and cooling it.

In this paper, we consider system that contains a mechanical cantilever and a NV center \cite{YIN2015}.
A second order magnetic gradient is applied to induce the coupling between
NV center and the two mechanical modes $a$ and $b$. The electron spin transition frequency of the NV center is tuned to match the mechanical modes frequencies difference.
The NV center is continuously resetted to the ground state by a laser. In this way, the NV center acts as an effective vacuum bath to
cool the mode $b$. An incoherent driving (thermal bath) is applied on the lower frequency mode $a$, and resulting a higher temperature of it.
The incoherent driving on mode $a$ also increases the effective coupling between mode $b$ and the NV center, and cools the mode $b$.
%Unlike the Ref. \cite{MAR2012}, where heat was directly injected to cavity mode and indirectly cooling the mechanical mode, here the mechanical
%resonator is cooled by  directly heating it.
We find that the quantum regime cooling is possible under the current experimental conditions.

The paper is orgnized as follows. In section II, we introduce the scheme of the second order magnetic field gradient
induced coupling between NV center and mechanical modes. In section III, we derived the analytical solution for the
system. In Section IV, the analytical results are verified by numerical simulation. In section V, a discussion and conclusion are given.

%%%%%%%%%%%%%%%%%%%%%%%%%%%%%%%%%%%%%%%%%%%%%%%%%%%%%%%%%%%%%%%%%%%%%%%%
%\section{setup for the system}
%%%%%%%%%%%%%%%%%%%%%%%%%%%%%%%%%%%%%%%%%%%%%%%%%%%%%%%%%%%%%%%%%%%%%%%%%
%\section{model of the system \label{II}}

\section{The scheme}
\begin{figure}
[hptb]
\begin{center}
\includegraphics[width=5cm] {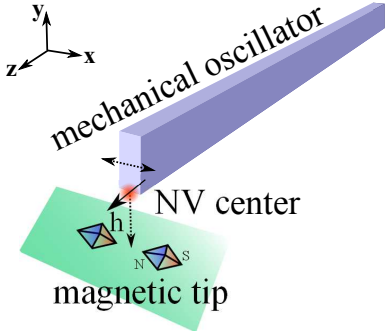}
\caption{A nano-diamond is fabricated as a thin cantilever which can oscillate along the $x$ axis, while the defect of an NV center is at the end. Two magnetic tips, which point along the $x$ axis and generate second order magnetic gradient along the $x$ axis, are placed below the cantilever.  The distance between the mid point of the connection lines of the tips and the NV center are around $50$ nm.
And an external static magnetic field $B_{ext}$ is added along the $z$ axis to induce Zeeman split in order to drive away the NV center spin $|+1\rangle$ state and keep the $|0\rangle$ and $|-1\rangle$ state closer.}%
\label{model}
%\vspace{0.2cm} %\hrule
\end{center}
\end{figure}

As shown in Fig.~\ref{model}, we consider a diamond cantilever resonator that contains a NV center at the end of it.
Two mechanical modes $a$ and $b$ are under consideration, with frequencies $\omega_a<\omega_b$.
Two spin states of the NV center electron are relevant here, $|-1\rangle$ and $|0\rangle$. By applying proper magnetic field, the energy split between the two states of the NV center, $\omega_z$, is equal to the energy difference of the two mechanical modes, $\Delta=\omega_b-\omega_a$.
Due to the  second order magnetic gradient $G_2= \partial^2 B/\partial x^2$, the two mechanical modes $a$ and $b$ both couple with the NV center.

Initially, both mechanical modes $a$ and $b$ are in the thermal states with the average thermal phonon number $\bar{n}_a$ and $\bar{n}_b$. The mode $a$ is driving by a thermal bath with average phonon number much higher than its environment. The NV center electron spin is initialized to the $|0\rangle$ state at the beginning.
The Hamiltonian of the whole system can be expressed as $H=H_0+H_1$, where $H_0$ is the non-interaction part and $H_1$ describes the interaction between the mechanical resonator and the NV center electron spin.

\begin{equation}
\label{HAM}
\begin{aligned}
H_0&=\frac{\omega_z}{2}\sigma_z+\omega_aa^{\dagger}a+\omega_bb^{\dagger}b\\
H_1&=\left(g_aa^{\dagger}a+g_bb^{\dagger}b+g_{ab}(a^{\dagger}b+ab^{\dagger})\right)\sigma_x
\end{aligned}
\end{equation}
where $g_a = g_s \mu_B G_2 x_a^2$, $g_b= g_s \mu_B G_2 x_b^2$, and $g_{ab} = g_s \mu_B G_2 x_a x_b$, and $x_a= \sqrt{\hbar/2m_a\omega_a}$ $(x_b=\sqrt{\hbar/2m_b\omega_b})$
is the zero field fluctuation for the mechanical mode $a$ $(b)$, $m_a$ ($m_b$) is the effective mass for mode $a$ ($b$). %(\textbf{Add some discussion on the present experimental parameters on the mechanical frequency and the effective mass, and get the value of $g_a$, $g_b$, and $g_{ab}$}.)
Experimentally, $\omega_z \sim 2\pi \times 10^7\ \rm{Hz}$, $g_a,g_b,g_{ab}\sim 2\pi \times 1\ \rm{Hz}$ can be realized (see Appendix).

Under the rotating wave approximation which shift the energy zero point by $\omega_a(a^{\dagger}a+b^{\dagger}b)$, we can change $H_0$ into
\begin{equation}
H_0'=\frac{\omega_z}{2}\sigma_z+\Delta b^{\dagger}b
\end{equation}
Unavoidably, the system will go through some intrinsic loss. The mechanical decay $\gamma_a$ and $\gamma_b$ is in the order of $2\pi\times1\ \rm{Hz}$ (see Appendix). The intrinsic decay rate $\Gamma_0$ for NV center is much less than $1$ Hz at low temperature. By adding initializing laser, the decay rate $\Gamma$ for the NV center electron spin would be much larger than the decay rates for the mechanical modes $\gamma_a$ and $\gamma_b$. The dephasing of NV center is neglected, as it could be eliminated by continuous dynamical decoupling \cite{Cai2012}. However, as we will use the same model to describe the decay of the spin and the vibration, here we should make sure that $\Gamma, \gamma_a, \gamma_b\ll\omega_z$.
As the evolution of the system contains decay and damping, it is convenient to describe it with quantum master equation~\cite{OPS}, which can be written out explicitly as
\begin{equation}
\label{RHO}
\begin{aligned}
\dot{\rho}&=-i[H,\rho]+(\mathcal{L}_a+\mathcal{L}_b+\mathcal{L}_s)\rho\\
\mathcal{L}_a&=(1+\bar{n}_a)\gamma_aD_a+\bar{n}_a\gamma_aD_{a^{\dagger}}\\
\mathcal{L}_b&=(1+\bar{n}_b)\gamma_bD_b+\bar{n}_b\gamma_bD_{b^{\dagger}}\\
\mathcal{L}_s&=\Gamma D_{\sigma_{-}}
\end{aligned}
\end{equation}
$D_x$ refers to the notation in Lindblad form, which is $D_x(\rho)=2x\rho x^{\dagger}-x^{\dagger}x\rho-\rho x^{\dagger}x$.
The Eq.\eqref{RHO} will be solved in both analytical and numerical methods. The results of the both methods will be compared and discussed.

\section{Analytical Solution}
As mentioned above, the phonon bath for mode $a$ is very large, so the average phonon number of mode $a$ is almost always equal to the average phonon number in the bath, that is, $\langle a^{\dagger}a \rangle=\bar{n}_a$. Moreover, as the phonon number of mode $b$ is much smaller than that of mode $a$, it is safe to drop the term $b^{\dagger}b$ in $H_1$. If we denote $\delta=a^{\dagger}a-\bar{n}_a$ as the phonon occupation number fluctuation of mode $a$, under the rotating wave picture, $H_1$ can be simplified to

\begin{equation}
H_1'=g_a\delta(\sigma_++\sigma_-)+g_{ab}(a^{\dagger}b\sigma_++ab^{\dagger}\sigma_-)
\end{equation}

For the convenience of the notation, here we rename $H_0'\rightarrow H_0$, $H_1'\rightarrow H_1$.

Using standard method of quantum master equation~\cite{QNO}, and after adiabatically eliminating the mode $a$ \cite{Yin2007,Yang2015}, 
we can get the reduced master equation for mode $b$ and the electron spin mode $s$ (see Appendix)

\begin{equation}
\label{RED}
\begin{aligned}
\dot{\rho}_{b,s}(t)=&-i[H_0,\rho_{b,s}(t)]+\mathcal{L}_{b}\rho_{b,s}(t)+\mathcal{L}_{s}\rho_{b,s}(t)\\
&+\mathcal{L}_1\rho_{b,s}(t)+\mathcal{L}_2\rho_{b,s}(t)
\end{aligned}
\end{equation}

where we have introduced two new Liouvillians for convenience

\begin{equation}
\label{LIOU}
\begin{aligned}
\mathcal{L}_{1}&=\frac{g_{ab}^2}{\gamma_a+\gamma_b+\Gamma}\left((1+\bar{n}_a)D_{b\sigma_+}+\bar{n}_aD_{b^{\dagger}\sigma_-}\right)\\
\mathcal{L}_{2}&=\frac{(2\gamma_a+\Gamma)g_a^2(\bar{n}_a^2+\bar{n}_a)}{\omega_z^2+(2\gamma_a+\Gamma)^2}(D_{\sigma_+}+D_{\sigma_-})
\end{aligned}
\end{equation}
Eq.~\eqref{LIOU} can be written in a more compact form if we define the effective average excitation number of the NV center electron spin as
\begin{equation}
n_s'=\frac{(2\gamma_a+\Gamma)g_a^2(\bar{n}_a^2+\bar{n}_a)}{\Gamma\left(\omega_z^2+(2\gamma_a+\Gamma)^2\right)}
\end{equation}

In fact, $n_s'$ is very close to zero because of the weak coupling requirement $(g_a\sqrt{\bar{n}_a},g_b\sqrt{\bar{n}_b},g_{ab}\sqrt{\bar{n}_a\bar{n}_b}\ll\omega_z)$, in other words, due to the large energy split $\omega_z$ and the relative large damping rate $\Gamma$. Then we can define an effective Liouvillian for the spin as well
$\mathcal{L}_s'=\mathcal{L}_s+\mathcal{L}_{2}=(1+n_s')\Gamma D_{\sigma_-}+n_s'\Gamma D_{\sigma_+}$.

Now the equation of motion for the reduced density matrix $\rho_{b,s}$ becomes
\begin{equation}
\label{DEN}
\dot{\rho}_{b,s}(t)=-i[H_0,\rho_{b,s}(t)]+\mathcal{L}_1\rho_{b,s}(t)+\mathcal{L}_b\rho_{b,s}(t)+\mathcal{L}_s'\rho_{b,s}(t)
\end{equation}
The corresponding adjoint equations for the phonon number operator $\hat{n}_b=b^{\dagger}b$ and excitation number operator $\hat{n}_s=\sigma_+\sigma_-$ can be developed then 
(see Appendix).
\begin{equation}
\label{NUMBER}
\begin{aligned}
\frac{d}{dt}\hat{n}_s=&\frac{2g_{ab}^2}{\gamma_a+\gamma_b+\Gamma}\Big[(-2\bar{n}_a-1)\hat{n}_b\hat{n}_s-\bar{n}_a\hat{n}_s\\&+(\bar{n}_a+1)\hat{n}_b\Big]
+2\Gamma\big[(-2n_s'-1)\hat{n}_s+n_s'\big]\\
\frac{d}{dt}\hat{n}_b=&\frac{2g_{ab}^2}{\gamma_a+\gamma_b+\Gamma}\Big[(2\bar{n}_a+1)\hat{n}_b\hat{n}_s+\bar{n}_a\hat{n}_s \\&-(\bar{n}_a+1)\hat{n}_b\Big]
+2\gamma_b(\bar{n}_b-\hat{n}_b)\\
\end{aligned}
\end{equation}

\begin{figure}[hbtp]
\begin{center}
\includegraphics[width=9cm] {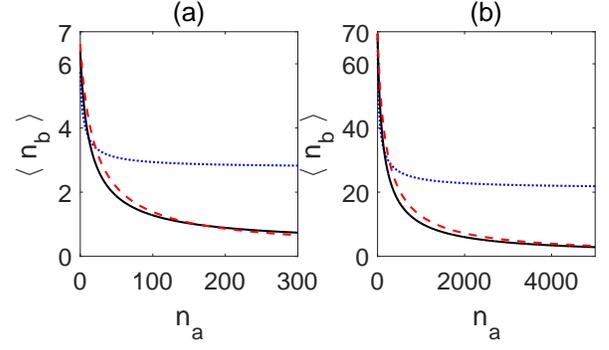}
\caption{$\langle \hat{n}_b \rangle$ as a function of $\bar{n}_a$. The parameters related to frequency are: $\omega_z/2\pi=2\times10^7$ Hz, $g_a/2\pi=3$ Hz, $g_b/2\pi=1$ Hz, $g_{ab}/2\pi=\sqrt{3}$ Hz, $\gamma_a/2\pi=\gamma_b/2\pi=1$ Hz.
%Two different environmental temperatures are shown in the figure.
(a) The initial temperature of mode $b$ is $10.8$ mK. The average phonon number is approximated by the Bose distribution $\bar{n}_b=1/(e^{\frac{\hbar\omega_b}{k_BT}}-1)=7$. Three values of $\Gamma/ 2\pi= 10, 30, 50$ Hz is corresponding to the blue dot line, black solid line and red dash line, respectively. $\Gamma/2\pi=30$ Hz and $50$ Hz can reach ground state cooling ($\langle \hat{n}_b \rangle<1$) if $\bar{n}_a>160$. This corresponds to an effective temperature $T_a=77.1\ \rm{mK}$. (b) The initial temperature of mode $b$ is $T=101.4\ \rm{mK}$. The thermal phonon number is $\bar{n}_b=70$. The blue dot, black solid and red dash lines are corresponding to $\Gamma/2\pi= 100, 300, 500$ Hz, respectively.
Ground state cooling can reach for $\Gamma/2\pi =300$ Hz and $\bar{n}_a>22500$ ($T_a=10.8\ \rm{K}$), or $\Gamma/2\pi=500$ Hz and $\bar{n}_a>19700$ ($T_a=9.5\ \rm{K}$)}%
\label{ANA}
%\vspace{0.2cm} %\hrule
\end{center}
\end{figure}

Taking the average on both sides respect to the quantum state of the system, and under average field approximation which refers to $\langle\hat{n}_b\hat{n}_s\rangle=\langle\hat{n}_b\rangle\langle\hat{n}_s\rangle$, the stationary state for the average phonon number of mode $b$ can be solved from the quadratic equation below after we drop the nonphysical solution.

\begin{equation}
\label{EQU}
A\langle \hat{n}_b \rangle^2+B\langle \hat{n}_b \rangle+C=0
\end{equation}
where
\begin{equation}
\label{COE}
\begin{aligned}
A=&\frac{-\gamma_bg_{ab}^2(2\bar{n}_a+1)}{\gamma_a+\gamma_b+\Gamma}\\
B=&\frac{g_{ab}^2}{\gamma_a+\gamma_b+\Gamma}(2\bar{n}_a\bar{n}_b\gamma_b+\bar{n}_b\gamma_b-\bar{n}_a\gamma_b\\
&-\Gamma\bar{n}_a-\Gamma n_s^{'}-\Gamma)-\gamma_b\Gamma(2n_s'+1)\\
C=&\frac{g_{ab}^2}{\gamma_a+\gamma_b+\Gamma}\bar{n}_a(\gamma_b\bar{n}_b+\Gamma n_s')+\gamma_b\bar{n}_b\Gamma(2n_s'+1)
\end{aligned}
\end{equation}
In Fig.~\ref{ANA}, the stationary phonon number of mode $b$ is plotted as a function of the average phonon number of mode $a$. Unlike the previous Ref.~\cite{MAR2012}, where there was an optimal heating bath temperature, here increasing heating bath temperature will always cooling the resonator, which is an outstanding property. As environment temperature increase, the quantum regime cooling $\langle \hat{n}_b \rangle<1$ can be realized by increasing both the mode $a$ temperature and NV center decay $\Gamma$.

\begin{figure}[hbtp]
\begin{center}
\includegraphics[width=7cm] {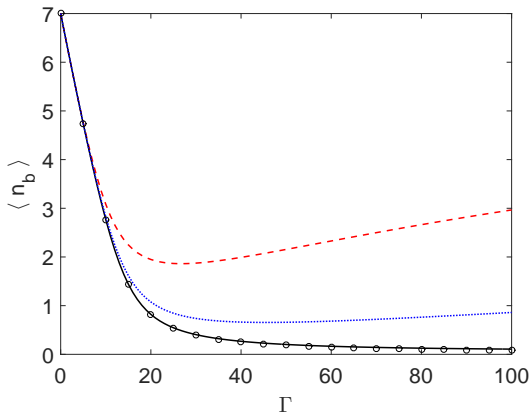}
\caption{$\langle \hat{n}_b \rangle$ as a function of $\Gamma$. Three values of $\bar{n}_a$, 50, 300, 10000, are corresponding to the red dash line, the blue dot line and the black solid line. Black circles display the Eq. \eqref{eq:nb1} in the limit of  $\bar{n}_a\rightarrow\infty$. The initial temperature of mode $b$ is $10.8\ \rm{mK}$ so $\bar{n}_b=7$.
%The value of $\Gamma/2\pi$ changes from $0$ to $100$ Hz.
Other parameters are same as Fig.~\ref{ANA}.}%
\label{ANA2}
%\vspace{0.2cm} %\hrule
\end{center}
\end{figure}

We can further look at how different values of $\Gamma$ can change the effect of cooling. The qualitative trend can be seen clearly from Fig.~\ref{ANA2}. For a given value of $\bar{n}_a$, when $\Gamma$ increases from a small value to a relative large value, there exists a certain $\Gamma$ for which the cooling effect is the largest.
%If we increase $\Gamma$ from 0 to a very large number (see Figure~\ref{ANA2}), the cooling effect increases firstly and then decreases.
Before the optimal cooling point, $\Gamma$ is not large compared with the effective coupling (e.g., $g_a\sqrt{\bar{n}_a}$), so increase $\Gamma$ will benefit the cooling process. However, after that, larger $\Gamma$ will result in smaller cooling rate. %term ($\Gamma$ is in the denominator of $\mathcal{L}_1$).
When $\bar{n}_a$ increases, the optimal $\Gamma$ will increase, too. In the limit of $\bar{n}_a\rightarrow \infty$, we can get a simple expression for $\langle \hat{n}_b \rangle$ under the weak coupling assumption
\begin{equation} \label{eq:nb}
\langle \hat{n}_b \rangle=\frac{1}{2}\left(\bar{n}_b-\frac{\gamma_b+\Gamma}{2\gamma_b}+\sqrt{(\bar{n}_b-\frac{\gamma_b+\Gamma}{2\gamma_b})^2+2\bar{n}_b}\right)
\end{equation}
As shown in Fig.~\ref{ANA2}, such an approximation is quite reasonable. In the limit of $\bar{n}_a\rightarrow \infty$, the condition for quantum regime cooling ($\langle \hat{n}_b \rangle<1$) is
\begin{equation}\label{eq:nb1}
\Gamma>3\gamma_b(\bar{n}_b-1),
\end{equation}
which can be fulfilled by tuning $\Gamma$ with laser.
These analytical results are based on rotating wave approximation (RWA). When thermal phonon number  $\sqrt{\bar{n}_a} \sim \Delta/g_{ab}$,
the RWA is no longer valid. From the parameters used in Fig. \ref{ANA}, RWA requires $\bar{n}_{ab} < 10^{12}$, which is obviously fulfilled in
Fig. \ref{ANA} and the following figures in this paper.

%%%%%%%%%%%%%%%%%%%%%%%%%%%%%%%%%%%%%%%%%%%%%%%%%%%%%%%%%%%%%%%%%%%%%%%
%\subsection{Numerical Simulation}
%%%%%%%%%%%%%%%%%%%%%%%%%%%%%%%%%%%%%%%%%%%%%%%%%%%%%%%%%%%%%%%%%%%%%%%%%%%
\section{Numerical Simulation}
Now we discuss the numerical simulation of the cooling process, and compare the result with the analytical ones above. The idea is simple. We expand the quantum operators and quantum states in the uncoupled representation, that is, the Kronecker tensor product of the occupation number representations for the three modes involved. The matrix forms of Eq.~\eqref{RHO} can then be calculated step by step with the classical four-order Runge-Kutta method. The average phonon number for mode $b$ is equal to $\mathrm{tr}(\rho N_b)$, where $N_b$ is the phonon number operator of mode $b$ in the whole Hilbert space for the system, namely, $N_b=I_s\otimes I_a\otimes(b^{\dagger}b)$. It should be pointed out, however, as the dimension of a matrix cannot be too large for a numerical simulation, the requirement $\bar{n}_a\gg\bar{n}_b$ is not able to be satisfied here. As a result, $H_1$ in Eq.~\eqref{HAM} should be inserted in the equation of motion Eq.~\eqref{RHO} instead of $H_1'$.

\begin{figure}[hbtp]
\begin{center}
\includegraphics[width=7cm] {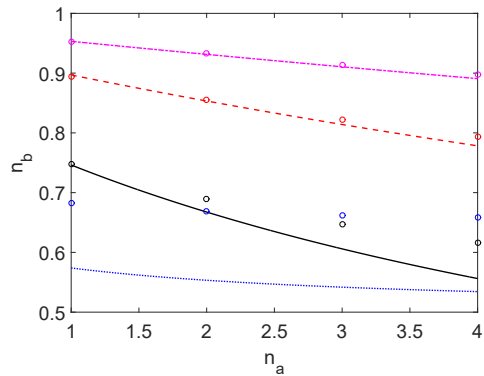}
\caption{(color online) Comparison of analytical and numerical results. Lines (circles) represent analytical (numerical) results. We choose four $\Gamma$: $2$ Hz, $15$ Hz, $50$ Hz, $120$ Hz, corresponding to the blue dot line (blue circles), black solid line (black circles), red dash line (red circles) and pink dash dot line (pink circles). The initial phonon number for mode $b$ is taken as $\bar{n}_b=1$. Other parameters are same as Fig.~\ref{ANA}.}%
%\vspace{0.2cm} %\hrule
\label{AN}
\end{center}
\end{figure}

We present several numerical results of the stationary phonon number $\langle \hat{n}_b \rangle$ with the results we get from Eq.~\eqref{EQU}, which is plotted in Fig.~\ref{AN}. As can be seen from the figure, when $\Gamma\gg\gamma_a,\gamma_b$ is satisfied, the analytical results match the numerical ones perfectly. But when $\gamma_a$ and $\gamma_b$ is comparable with $\Gamma$, the two results no longer agree with each other quantitatively. However, it is quite reasonable, because in this case several approximations for the analytical results are not valid any more. When $\Gamma$ is not very large, the term $\mathcal{L}_s=\Gamma D_{\sigma_-}$ does not dominate over $H$ in Eq.~\eqref{RHO}. As $\bar{n}_a$ is also at the same scale with $\bar{n}_b$ in this case, now neglecting the back action on mode $a$ is not a good approximation. When mode $b$ is being cooled, mode $a$ is being heated unavoidably, thus lowering the effect of cooling. The analytical results which neglect the back action thus exaggerate the cooling effect. This trend is shown clearly from the figure. But the analytical conclusions are still valid qualitatively. For instance, if we fix $\Gamma$, the larger $\bar{n}_a$ is, the larger the cooling effect will be. And if we fix $\bar{n}_a$, there exists a $\bar{n}_a$-dependent value of $\Gamma$ where the cooling effect is maximum.

%A more detailed discussion of the numerical simulation can be find in Appendix~\ref{APP3}.

%%%%%%%%%%%%%%%%%%%%%%%%%%%%%%%%%%%%%%%%%%%%%%%%%%%%%%%%%%%%%%%%%%%%%%%%
%\section{conclusion and discussion\label{SEC}}
%%%%%%%%%%%%%%%%%%%%%%%%%%%%%%%%%%%%%%%%%%%%%%%%%%%%%%%%%%%%%%%%%%%%%%%%%
\section{Discussion and conclusion}
From the above derivation and calculation, it is clear the higher frequency mechanical mode will be significantly cooled by heating the lower frequency mechanical mode. In fact, this trend can be seen directly from the Eq.~\eqref{DEN},
where $\mathcal{L}_1$ acts as the cooling term. As the electron spin is much more likely to be in the spin state $|0\rangle$ due to the continuous initialization, the effect of $\sigma_+$ should be much larger than that of $\sigma_-$. So $D_{b\sigma_+}$ is the dominant operator for evolution, and the mode $b$ will be cooled.  This can be understood that mode $a$ facilitate the energy transfer between mode $b$ and the spin. Quantized energy flows from the vibration mode to the NV center. Soon after that the excitation in the NV center electron spin decays into the environment through light. As this process is repeated, the heat of the mechanical mode $b$ is gradually removed.

In conclusion, we studied the second order magnetic gradient induced coupling between NV center and the mechanical modes. We proposed to drive one mechanical resonator mode with thermal bath to cool the other mode of the resonator. We solved the motion of the system analytically under the weak coupling assumption. It is found that under rotating wave approximation final thermal phonon number of the resonator approach to a minimum as the thermal bath temperature increases. The quantum regime cooling is feasible under the current experimental conditions. We numerically solved the equations, and found that the results fitted very well with the analytic results. The cooling by heating phenomena does not violate the laws of thermodynamics. This scheme can be understood as a thermal machine or a heat engine \cite{Zhang2014,Mari2015,Mitchison}. The mode $a$ is an ``engine'', and the NV center electron spin takes the role of a ``condenser''. This work opens up the possibility that cooling the mechanical mode in solid systems by injecting the thermal phonon, which usually heats the system.

\section{Acknowledgements}
Z.Q.Y. is supported by National Natural Science Foundation of China Grant 61435007. W.L.Y. is supported by National Natural Science Foundation of China Grant 11574353 and 11274351. P.H and J.D. is supported by 973 Program (Grant No. 2013CB921800), the National Natural Science Foundation of China (Grant No. 11227901), and the CAS (Grant No. XDB01030400).

\appendix
\section{Experimentally realized paramters}
%%%%%%%%%%%%%%%%%%%%%%%%%%%%%%%%%%%%%%%%%%%%%%%%%%%%%%%%%%%%%%%%%%%

In this section we discuss the values of parameters $\omega_z$, $\omega_a$, $\omega_b$, $g_a$, $g_b$, $g_{ab}$, $\gamma_a$, $\gamma_b$, $\Gamma$ that can be realized in experiments.

As in former experiments (e.g.~\cite{RAB2009} for a Si cantilever), the size of the resonator can be taken as around $3\rm{\mu m}\times0.05\rm{\mu m}\times0.05\rm{\mu m}$. Considering that the density of diamond is around 1.5 times the density of silicon~\cite{Tao2014}, and the fact that the effective mass of a solid and normal shaped oscillator doing simple oscillation is of the same order as its mass, we can estimate that $m_a=m_b=5\times10^{-18}$ kg. The fundamental frequency of such a mechanical resonator is around 10 MHz~\cite{RAB2009}. Although the zero-field splitting of the NV center is around 2.88 GHz~\cite{Wrachtrup2006}, after an external magnetic field is exerted, the separation between $|0\rangle$ and $|-1\rangle$ states can be reduced to the order of tens of MHz. As a result, it is possible to fulfill the requirement $\omega_b-\omega_a=\omega_z$. Here we choose $\omega_a/2\pi=10\ \rm{MHz}$, $\omega_b/ 2\pi =30\ \rm{MHz}$ for convenience. The quality factor of the resonator can reach up to $Q=10^7$~\cite{Tao2014}. Second-order magnetic gradient $G_2$ can realize several times of $10^{13}\ \mathrm{T\cdot m^{-2}}$ in experiment~\cite{Mamin2007}. By using finite element simulation, we got that the maximum $G_2$ could be around $5\times 10^{14}\ \rm{T\cdot m^{-2}}$, if the distance between NV center and the magnetic particles is $50$ nm.

If $G_2$ is set to be $5\times10^{14}\ \rm{T\cdot m^{-2}}$, we will have $g_a/2\pi=3\ \rm{Hz}$, $g_b/2\pi=1\ \rm{Hz}$, $g_{ab}/2\pi=1.73\ \rm{Hz}$. The decay rates of the mechanical resonator can be estimated by $\gamma_a=\gamma_b=\omega_a/Q=2\pi \times 1\ \rm{Hz}$. We need to tune the decay rate of the NV center larger enough than the mechanical decay rate to reach ground state cooling, for instance, $\Gamma$ can equals to $2\pi\times 120\ \rm{Hz}$.

%%%%%%%%%%%%%%%%%%%%%%%%%%%%%%%%%%%%%%%%%%%%%%%%%%%%%%%%%%%%%%%%%%%
\section{Derivation of the Reduced Master Equation}\label{APP1}
%%%%%%%%%%%%%%%%%%%%%%%%%%%%%%%%%%%%%%%%%%%%%%%%%%%%%%%%%%%%%%%%%%%

Here we provide a derivation of the reduced master equation Eq.~\eqref{RED} in main text.
The reduced density matrix of mode $b$ and mode $s$ after mode $a$ is eliminated is

\begin{equation}
\rho_{b,s}=\mathrm{tr}_a(\rho)
\end{equation}

We can adiabatically eliminate mode $a$ to get the evolution equation for the reduced density matrix \cite{QNO}

\begin{equation}
\begin{aligned}
\dot{\rho}_{b,s}(t)=&-i[H_0,\rho_{b,s}(t)]+\mathcal{L}_{b}\rho_{b,s}(t)+\mathcal{L}_{s}\rho_{b,s}(t)\\
&+\mathrm{tr}_a\mathcal{L}_{int}\int_0^{\infty}dt^{'}e^{\mathcal{L}_{free}t^{'}}
\mathcal{L}_{int}\rho_{b,s}(t-t^{'})\otimes\rho_a\\
=&-i[H_0,\rho_{b,s}(t)]+\mathcal{L}_{b}\rho_{b,s}(t)+\mathcal{L}_{s}\rho_{b,s}(t)\\
&-\mathrm{tr}_a\left[H_1,\int_0^{\infty}dt^{'}e^{\mathcal{L}_{free}t^{'}}[H_1,\rho_{b,s}(t-t^{'})\otimes\rho_a]\right]
\end{aligned}
\end{equation}

where we denote

\begin{equation}
\begin{aligned}
\mathcal{L}_{int}&=-i[H_1,\bullet]\\
\mathcal{L}_{free}&=-i[H_0,\bullet]+\mathcal{L}_a+\mathcal{L}_b+\mathcal{L}_s
\end{aligned}
\end{equation}

We expand $g_a$, $g_b$ and $g_{ab}$ to the second order and can get

\begin{equation}
\label{EOM}
\begin{aligned}
\dot{\rho}_{b,s}(t)=&-i[H_0,\rho_{b,s}(t)]+\mathcal{L}_{b}\rho_{b,s}(t)+\mathcal{L}_{s}\rho_{b,s}(t)\\
&-\mathrm{tr}_a\left[H_1,\int_0^{\infty}dt^{'}[e^{\mathcal{L}_{free}^{\dagger}t^{'}}(H_1),\rho_{b,s}(t)\otimes\rho_a]\right]
\end{aligned}
\end{equation}

Once we expand the commutation relations in Eq.~\eqref{EOM}, we can directly calculated it term by term.

\begin{equation}
\label{DOTRHO}
\begin{aligned}
\dot{\rho}_{b,s}(t)=&-i[H_0,\rho_{b,s}(t)]+\mathcal{L}_{b}\rho_{b,s}(t)+\mathcal{L}_{s}\rho_{b,s}(t)\\
&-\mathrm{tr}_a\bigg(\int_0^{\infty}dt^{'}\Big(H_1e^{\mathcal{L}_{free}^{\dagger}t^{'}}(H_1)\rho_{b,s}(t)\otimes\rho_a\\
&-H_1\rho_{b,s}(t)\otimes\rho_ae^{\mathcal{L}_{free}^{\dagger}t^{'}}(H_1)\\
&-e^{\mathcal{L}_{free}^{\dagger}t^{'}}(H_1)\rho_{b,s}(t)\otimes\rho_aH_1\\
&+\rho_{b,s}(t)\otimes\rho_ae^{\mathcal{L}_{free}^{\dagger}t^{'}}(H_1)H_1 \Big) \bigg)
\end{aligned}
\end{equation}

What we need first is the term $e^{\mathcal{L}_{free}^{\dagger}t^{'}}(H_1)$. It is easy to get from direct calculation.

\begin{equation}
\begin{aligned}
\mathcal{L}_{free}^{\dagger}(a)=&(1+\bar{n}_a)\gamma_aD_a^{\dagger}(a)+\bar{n}_a\gamma_aD_{a^{\dagger}}^{\dagger}(a)\\
=&(1+\bar{n}_a)\gamma_a(2a^{\dagger}a^2-a^{\dagger}a^2-aa^{\dagger}a)\\
&+\bar{n}_a\gamma_a(2a^2a^{\dagger}-aa^{\dagger}a-a^2a^{\dagger})\\
=&-\gamma_aa
\end{aligned}
\end{equation}

Similarly, $\mathcal{L}_{free}^{\dagger}(a^{\dagger})=-\gamma_aa^{\dagger}$. From these we have $e^{\mathcal{L}_{free}^{\dagger}t^{'}}a=e^{-\gamma_at^{'}}a$, $e^{\mathcal{L}_{free}^{\dagger}t^{'}}a^{\dagger}=e^{-\gamma_at^{'}}a^{\dagger}$.

The following results are got using exactly the same method: %$e^{\mathcal{L}_{free}^{\dagger}t^{'}}\delta=e^{-2\gamma_at^{'}}\delta$, $e^{\mathcal{L}_{free}^{\dagger}t^{'}}b=e^{-(\gamma_b+i\omega_z)t^{'}}b$, $e^{\mathcal{L}_{free}^{\dagger}t^{'}}b^{\dagger}=e^{-(\gamma_b-i\omega_z)t^{'}}b$, $e^{\mathcal{L}_{free}^{\dagger}t^{'}}\sigma_+=e^{(i\omega_z-\Gamma)t^{'}}\sigma_+$, $e^{\mathcal{L}_{free}^{\dagger}t^{'}}\sigma_-=e^{(-i\omega_z-\Gamma)t^{'}}\sigma_-$.

\begin{equation}
\begin{aligned}
&e^{\mathcal{L}_{free}^{\dagger}t^{'}}\delta=e^{-2\gamma_at^{'}}\delta,\ e^{\mathcal{L}_{free}^{\dagger}t^{'}}b=e^{-(\gamma_b+i\omega_z)t^{'}}b\\
&e^{\mathcal{L}_{free}^{\dagger}t^{'}}b^{\dagger}=e^{-(\gamma_b-i\omega_z)t^{'}}b,\ e^{\mathcal{L}_{free}^{\dagger}t^{'}}\sigma_+=e^{(i\omega_z-\Gamma)t^{'}}\sigma_+\\
&e^{\mathcal{L}_{free}^{\dagger}t^{'}}\sigma_-=e^{(-i\omega_z-\Gamma)t^{'}}\sigma_-
\end{aligned}
\end{equation}

$e^{\mathcal{L}_{free}^{\dagger}t^{'}}(H_1)$ will then be calculated with the results above in mind.

\begin{equation}
\begin{aligned}
&e^{\mathcal{L}_{free}^{\dagger}t^{'}}(H_1)\\
=&g_a\Big((e^{\mathcal{L}_{free}^{\dagger}t^{'}}\delta)(e^{\mathcal{L}_{free}^{\dagger}t^{'}}\sigma_+)
+(e^{\mathcal{L}_{free}^{\dagger}t^{'}}\delta)(e^{\mathcal{L}_{free}^{\dagger}t^{'}}\sigma_-)\Big)\\
&+g_{ab}\Big((e^{\mathcal{L}_{free}^{\dagger}t^{'}}a^{\dagger})(e^{\mathcal{L}_{free}^{\dagger}t^{'}}b)(e^{\mathcal{L}_{free}^{\dagger}t^{'}}\sigma_+)\\
&+(e^{\mathcal{L}_{free}^{\dagger}t^{'}}a)(e^{\mathcal{L}_{free}^{\dagger}t^{'}}b^{\dagger})(e^{\mathcal{L}_{free}^{\dagger}t^{'}}\sigma_-)\Big)\\
=&g_a\delta e^{-(2\gamma_a+\Gamma)t^{'}}(\sigma_+e^{i\omega_zt^{'}}+\sigma_-e^{-i\omega_zt^{'}})\\
&+g_{ab}e^{-(\gamma_a+\gamma_b+\Gamma)t^{'}}(a^{\dagger}b\sigma_++ab^{\dagger}\sigma_-)
\end{aligned}
\end{equation}

It deserves notice that here we only keep the slow-time-varying term, so we neglect $a^{\dagger2}$ and $a^2$ above. Now we come back to Eq.~\eqref{DOTRHO}, which can be simplified term by term. As $\mathrm{tr}_a$ means taking the average on $a$, the trace of odd multiples of $a$ and $a^{\dagger}$ equals to zero. Terms containing $i\omega_z$ are neglected here because their effect is just changing the energy zero point.

\begin{equation}
\label{CAL1}
\begin{aligned}
&\mathrm{tr}_a\Big(\int_0^{\infty}dt^{'}H_1e^{\mathcal{L}_{free}^{\dagger}t^{'}}(H_1)\rho_{b,s}(t)\otimes\rho_a\Big)\\
=&g_a^2(\bar{n}_a^2+\bar{n}_a)\frac{2\gamma_a+\Gamma}{\omega_z^2+(2\gamma_a+\Gamma)^2}(\sigma_+\sigma_-+\sigma_-\sigma_+)\rho_{b,s}(t)\\
&+\frac{g_{ab}^2}{\gamma_a+\gamma_b+\Gamma}\Big((\bar{n}_a+1)b^{\dagger}\sigma_-b\sigma_++\bar{n}_ab\sigma_+b^{\dagger}\sigma_-\Big)\rho_{b,s}(t)
\end{aligned}
\end{equation}

\begin{equation}
\label{CAL2}
\begin{aligned}
&\mathrm{tr}_a\Big(\int_0^{\infty}dt^{'}H_1\rho_{b,s}(t)\otimes\rho_a e^{\mathcal{L}_{free}^{\dagger}t^{'}}(H_1)\Big)\\
=&g_a^2(\bar{n}_a^2+\bar{n}_a)\frac{2\gamma_a+\Gamma}{\omega_z^2+(2\gamma_a+\Gamma)^2}(\sigma_+\rho_{b,s}(t)\sigma_-+\sigma_-\rho_{b,s}(t)\sigma_+)\\
&+\frac{g_{ab}^2}{\gamma_a+\gamma_b+\Gamma}\Big((\bar{n}_a+1)b\sigma_+\rho_{b,s}(t)b^{\dagger}\sigma_-+\bar{n}_ab^{\dagger}\sigma_-\rho_{b,s}(t)b\sigma_+\Big)
\end{aligned}
\end{equation}

\begin{equation}
\label{CAL3}
\begin{aligned}
&\mathrm{tr}_a\Big(\int_0^{\infty}dt^{'}e^{\mathcal{L}_{free}^{\dagger}t^{'}}(H_1)\rho_{b,s}(t)\otimes\rho_a H_1\Big)\\
=&g_a^2(\bar{n}_a^2+\bar{n}_a)\frac{2\gamma_a+\Gamma}{\omega_z^2+(2\gamma_a+\Gamma)^2}(\sigma_+\rho_{b,s}(t)\sigma_-+\sigma_-\rho_{b,s}(t)\sigma_+)\\
&+\frac{g_{ab}^2}{\gamma_a+\gamma_b+\Gamma}\Big((\bar{n}_a+1)b\sigma_+\rho_{b,s}(t)b^{\dagger}\sigma_-+\bar{n}_ab^{\dagger}\sigma_-\rho_{b,s}(t)b\sigma_+\Big)
\end{aligned}
\end{equation}

\begin{equation}
\label{CAL4}
\begin{aligned}
&\mathrm{tr}_a\Big(\int_0^{\infty}dt^{'}\rho_{b,s}(t)\otimes\rho_ae^{\mathcal{L}_{free}^{\dagger}t^{'}}(H_1)H_1\Big)\\
=&g_a^2(\bar{n}_a^2+\bar{n}_a)\frac{2\gamma_a+\Gamma}{\omega_z^2+(2\gamma_a+\Gamma)^2}\rho_{b,s}(t)(\sigma_+\sigma_-+\sigma_-\sigma_+)\\
&+\frac{g_{ab}^2}{\gamma_a+\gamma_b+\Gamma}\rho_{b,s}(t)\Big((\bar{n}_a+1)b^{\dagger}\sigma_-b\sigma_++\bar{n}_ab\sigma_+b^{\dagger}\sigma_-\Big)
\end{aligned}
\end{equation}

After we insert Eq.~\eqref{CAL1},~\eqref{CAL2},~\eqref{CAL3},~\eqref{CAL4} into Eq.~\eqref{DOTRHO}, we will get Eq.~\eqref{RED} in the main text.

%%%%%%%%%%%%%%%%%%%%%%%%%%%%%%%%%%%%%%%%%%%%%%%%%%%%%%%%%%%%%%%%%%%
\section{Derivation of the Adjoint Equations for the Number Operators}\label{APP2}
%%%%%%%%%%%%%%%%%%%%%%%%%%%%%%%%%%%%%%%%%%%%%%%%%%%%%%%%%%%%%%%%%%%

In this part, the equations of motion for the number operators $\hat{n}_b$ and $\hat{n}_s$, Eq.~\eqref{NUMBER} in main text, is derived.

The adjoint equations for Eq.~\eqref{NUMBER} in main text are

\begin{equation}
\label{ADJ}
\begin{aligned}
\frac{d}{dt}\hat{n}_s=&(\mathcal{L}_1^{\dagger}+\mathcal{L}_s^{'\dagger})\hat{n}_s\\
=&\frac{g_{ab}^2}{\gamma_a+\gamma_b+\Gamma}\left((1+\bar{n}_a)D_{b\sigma_+}^{\dagger}+\bar{n}_aD_{b^{\dagger}\sigma_-}^{\dagger}\right)\hat{n}_s\\
&+\left((1+n_s^{'})\Gamma D_{\sigma_-}^{\dagger}+n_s^{'}\Gamma D_{\sigma_+}^{\dagger}\right)\hat{n}_s\\
\frac{d}{dt}\hat{n}_b=&(\mathcal{L}_1^{\dagger}+\mathcal{L}_b^{\dagger})\hat{n}_b\\
=&\frac{g_{ab}^2}{\gamma_a+\gamma_b+\Gamma}\left((1+\bar{n}_a)D_{b\sigma_+}^{\dagger}+\bar{n}_aD_{b^{\dagger}\sigma_-}^{\dagger}\right)\hat{n}_b\\
&+\left((1+\bar{n}_b)\gamma_b D_{b}^{\dagger}+\bar{n}_b\gamma_b D_{b^{\dagger}}^{\dagger}\right)\hat{n}_b
\end{aligned}
\end{equation}

From simple calculations, we have

\begin{equation}
\label{ADJ2}
\begin{aligned}
&D_{b\sigma_+}^{\dagger}(\hat{n}_s)=2(1-\hat{n}_s)\hat{n}_b,\ D_{b^{\dagger}\sigma_-}^{\dagger}(\hat{n}_s)=-2\hat{n}_s(1+\hat{n}_b)\\
&D_{\sigma_+}^{\dagger}(\hat{n}_s)=2(1-\hat{n}_s),\ D_{\sigma_-}^{\dagger}(\hat{n}_s)=-2\hat{n}_s\\
&D_{b\sigma_+}^{\dagger}(\hat{n}_b)=-2(1-\hat{n}_s)\hat{n}_b,\ D_{b^{\dagger}\sigma_-}^{\dagger}(\hat{n}_b)=2\hat{n}_s(1+\hat{n}_b)\\
&D_{b}^{\dagger}(\hat{n}_b)=-2\hat{n}_b,\ D_{b^{\dagger}}^{\dagger}(\hat{n}_b)=2(1+\hat{n}_b)\\
\end{aligned}
\end{equation}

After Eq.~\eqref{ADJ2} is inserted into Eq.~\eqref{ADJ}, it will be simplified to Eq.~\eqref{NUMBER} in the main text.

%%%%%%%%%%%%%%%%%%%%%%%%%%%%%%%%%%%%%%%%%%%%%%%%%%%%%%%%%%%%%%%%%%%
\section{Detail of the Numerical Simulation}\label{APP3}
%%%%%%%%%%%%%%%%%%%%%%%%%%%%%%%%%%%%%%%%%%%%%%%%%%%%%%%%%%%%%%%%%%%

In this section, we discuss in detail the numerical simulation.
For Fig.~\ref{AN} in main text, we control the accuracy so that all numbers
are accurate until the order of 0.001. The evolution time is $t=3$, after which
the state is at the stationary state up to the accuracy. As discussed in the Appendix before,
we can renormalize the frequency parameters as $g_{ab}=\sqrt{3}$, $g_b=1$, $g_a=3$, $\gamma_a=\gamma_b=1$.
As we want to cover a wide range of results, four values renormalized $\Gamma$ are chosen: 2, 15, 50 and 120.
Small values for the phonon numbers $\bar{n}_a$ and $\bar{n}_b$ should be chosen due to the calculation cost.

\begin{figure}[tbp]
\begin{center}
\includegraphics[width=6cm] {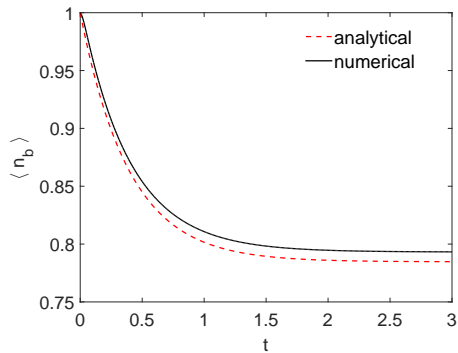}
\caption{The time evolution of the phonon number of mode $b$. Here $\Gamma=50$. Initial phonon numbers are: $\bar{n}_a=4$, $\bar{n}_b=1$. The matrix dimension of the operators and states for mode $a$ is taken to be $60\times60$, for mode $b$ it is $15\times15$. Other parameters are same as Fig.~\ref{ANA} in main text. For the analytical solution of the differential equation Eq.~\eqref{NUMBER} in main text, standard solver ode45 in Matlab is used.}%
\label{NUM}
%\vspace{0.2cm} %\hrule
\end{center}
\end{figure}

Here we always choose $\bar{n}_b=1$ as the initial condition. As to $\bar{n}_a$, we choose $\bar{n}_a=1,2,3,4$.
The dimensions of the matrices in the Hilbert space are chosen to balance the accuracy and the cost of calculation resource.
For the Hilbert space where the initial phonon number equals to 1, the dimension is chosen as $15\times15$. For the Hilbert
space where the initial phonon number equals to 2, the dimension is chosen as $27\times27$. For the Hilbert space where the initial phonon number equals to 3, the dimension is chosen as $40\times40$. For the Hilbert space where the initial phonon number equals to 4, the dimension is chosen as $60\times60$. Obviously, the dimension of the Hilbert space for mode $s$ is always $2\times2$. The step length for the Runge-Kutta method is chosen to be $0.01/50$.

Fig.~\ref{AN} in the main text shows the results of only the $\omega_z=500$ case, even if for the example in the Appendix, $\omega_z$ should be renormalized as $\omega_z=2\times10^7$. In fact, we have tried that different $\omega_z$ will not influence the result of cooling, e.g., $\omega_z=250,\ 500,\ 750$ all give out same final phonon number for mode $b$. So we are safe to use smaller $\omega_z$ here considering the stability of the simulation. This deduction also agrees with the analytical result because Eq.~\eqref{EQU} in main text does not contain $\omega_z$ as long as $\omega_z$ is large enough for $n_s^{'}\approx0$.

The time evolution of one set of parameters from numerical simulation is shown in Fig.~\ref{NUM}. The phonon occupation number $\langle \hat{n}_b \rangle$ gradually goes down as the system evolves towards the stationary state. The cooling effect indeed happens. Also shown in the figure is the time evolution starting from the analytical result Eq.~\eqref{NUMBER}. The two lines are similar with the cooling effect of the analytical one a little bit more obvious, which also supports our findings.

%\newpage

\end{document}